\documentclass{article}

\usepackage{microtype}
\usepackage{graphicx}
\usepackage{subfigure}
\usepackage{booktabs} 

\usepackage{hyperref}

\usepackage[accepted]{icml2024}

\usepackage{amsmath}
\usepackage{tabularx}
\usepackage{tabulary}
\usepackage{amssymb}
\usepackage{mathtools}
\usepackage{amsthm}
\usepackage{booktabs}
\usepackage{authblk}

\usepackage[capitalize,noabbrev]{cleveref}

\usepackage{amsmath,amsfonts,bm}

\def\eqref#1{equation~\ref{#1}}

\def\1{\bm{1}}

\def\mA{{\bm{A}}}

\def\mG{{\bm{G}}}

\def\mU{{\bm{U}}}
\def\mV{{\bm{V}}}

\DeclareMathAlphabet{\mathsfit}{\encodingdefault}{\sfdefault}{m}{sl}
\SetMathAlphabet{\mathsfit}{bold}{\encodingdefault}{\sfdefault}{bx}{n}

\def\gG{{\mathcal{G}}}

\DeclareMathOperator*{\argmin}{arg\,min}

\theoremstyle{plain}

\theoremstyle{definition}

\theoremstyle{remark}

\usepackage[textsize=tiny]{todonotes}

\newcommand{\squishlist}{
  \begin{list}{$\bullet$}
    { \setlength{\itemsep}{0pt}      \setlength{\parsep}{0pt}
      \setlength{\topsep}{0pt}       \setlength{\partopsep}{0pt}
      \setlength{\leftmargin}{1em} \setlength{\labelwidth}{1em}
      \setlength{\labelsep}{0.5em} } }
\newcommand{\squishlisttwo}{
  \begin{list}{$\bullet$}
    { \setlength{\itemsep}{0pt}    \setlength{\parsep}{0pt}
      \setlength{\topsep}{0pt}     \setlength{\partopsep}{0pt}
      \setlength{\leftmargin}{2em} \setlength{\labelwidth}{1.5em}
      \setlength{\labelsep}{0.5em} } }
\newcommand{\squishlistend}{
    \end{list}  }

\icmltitlerunning{Submission and Formatting Instructions for ICML 2024}

\begin{document}

\title{Dynamic Neural Control Flow Execution: An Agent-Based Deep Equilibrium Approach for Binary Vulnerability Detection}



\author[1]{Litao Li}
\author[1]{Steven~H.~H.~Ding}
\author[2]{Andrew~Walenstein}
\author[3]{Philippe~Charland}
\author[4]{Benjamin~C.~M.~Fung}
\affil[1]{L1NNA Lab, School of Computing, Queen's University, Canada}
\affil[2]{BlackBerry Ltd., Canada}
\affil[3]{Mission Critical Cyber Security Section, Defence R\&D Canada}
\affil[4]{Data Mining and Security (DMaS) Lab, McGill University, Canada}

\maketitle

\begin{abstract}
  Software vulnerabilities are a challenge in cybersecurity. Manual security patches are often difficult and slow to be deployed, while new vulnerabilities are created. Binary code vulnerability detection is less studied and more complex compared to source code, and this has important practical implications. Deep learning has become an efficient and powerful tool in the security domain, where it provides end-to-end and accurate prediction. Modern deep learning approaches learn the program semantics through sequence and graph neural networks, using various intermediate representation of programs, such as abstract syntax trees (AST) or control flow graphs (CFG). Due to the complex nature of program execution, the output of an execution depends on the many program states and inputs. Also, a CFG generated from static analysis can be an overestimation of the true program flow. Moreover, the size of programs often does not allow a graph neural network with fixed layers to aggregate global information. To address these issues, we propose DeepEXE, an agent-based implicit neural network that mimics the execution path of a program. We use reinforcement learning to enhance the branching decision at every program state transition and create a dynamic environment to learn the dependency between a vulnerability and certain program states. An implicitly defined neural network enables nearly infinite state transitions until convergence, which captures the structural information at a higher level. The experiments are conducted on two semi-synthetic and two real-world datasets. We show that DeepEXE is an accurate and efficient method and outperforms the state-of-the-art vulnerability detection methods.
\end{abstract}




\section{Introduction}
Software vulnerabilities have been an ongoing challenge in the cybersecurity domain. It is an inevitable problem, as the scale of software grows in complexity. Many malicious cyber attacks exploit vulnerabilities within systems and can cause tremendous economical and security damages. Often, the security analysts cannot even patch vulnerabilities fast enough, as new ones are created~\cite{alexopoulos2020tip, farris2018vulcon}. Common Vulnerability Exposures (CVE) show that the total number of vulnerabilities more than doubled from 2016 to 2017 and it continued to increase throughout the recent years\footnote{\href{https://www.cvedetails.com/}{Statistics on Common Vulnerabilities and Exposures (CVE) Details}}. Many traditional static and dynamic analysis methods are manually expensive and inefficient. This motivates automated and end-to-end approaches, such as neural networks.

Vulnerabilities can be detected at either the source code or binary code level. Source code provides much more meaningful semantics, syntax, and structures, which in turn help both analysts and machine learning models to track vulnerabilities. Existing methods at the source code level are accurate and capable of finding complex vulnerabilities~\cite{li2018vuldeepecker,harer2018automated}. For binary code, as much information is lost during the compilation process, it is much harder to detect vulnerabilities. Moreover, the absence of the original source is a practical problem under many circumstances, such as third-party or off-the-shelf programs. Binary code is best analyzed as assembly code, a form of intermediate representation that provides analysts readable content. Assembly code contains instructions that provide some semantics and structures of the program. In this paper, we are only interested in binary code vulnerability detection, as it is still a prevalent challenge in the security domain.

Deep learning methods aim to learn the latent representation of a piece of binary code for classification. Existing works for binary code learning can be categorized into two main streams. The first approach focuses on text-based representation learning to extract the token semantics. The instructions are broken down and embedded into vectors through some unsupervised learning such as Word2Vec~\cite{mikolov2013efficient}, then these vectors are fed into a sequential deep learning model for classification. \textbf{Instruction2Vec}~\cite{b10}, \textbf{HAN-BSVD}~\cite{yan2021han}, and \textbf{BVDetector}~\cite{tian2020bvdetector} all use this semantic-based approach for detection. The second method involves collecting and aggregating structural information at a higher level. Usually, CFGs are parsed from the assembly code basic blocks, which create dependencies between different blocks of code. This is crucial in vulnerability detection, since programs are complex and hierarchical, and vulnerabilities are often triggered in specific program states. Using only the semantics of instruction tokens are often insufficient. \textbf{Gemini}~\cite{xu2017neural}, \textbf{Diff}~\cite{liu2018alphadiff}, \textbf{Order}~\cite{yu2020order}, \textbf{InnerEye}~\cite{zuo2018neural}, and \textbf{BinDeep}~\cite{tian2021bindeep} all use graph-based methods for binary code structure embedding. 

Unfortunately, there are major drawbacks to either approach that can hinder the performance or scalability of the model. The more obvious disadvantage is the scalability when large programs are present. Semantic-based approaches usually introduce a maximum input length, in order to prevent vanishing gradient, especially for large and deep sequence models. Structure-based approaches perform graph neural network (GNN) for aggregating node information. The number of layers dictates the receptive field of the model by performing $k$-hop message passing, thus limiting the amount of global information that can be learned. Both of them need to carefully manage the memory footprint during training.
The other drawback is the absence of modelling how programs naturally run. Unlike natural language, programs are executed dynamically. The state of a program can be different, depending on the input and its previous states. By using fixed graph learning techniques, the dynamic nature of the program structure is difficult to capture and thus lead to undesired performance.

Given assembly code, one has to respectively find a program execution path that can potentially yield the same final program state. In general, a sound and complete static analysis method generates a representation of the code (i.e., CFG) with overestimation. This means paths created in a graph can potentially never execute. Therefore, learning the topological information solely from the default CFG can be inaccurate and result in a false execution path. Ideally, symbolic execution~\cite{baldoni2018survey, king1976symbolic} is one of the formal methods that enable one to compare and verify all the possible paths through equivalence checking. However, its applicability is limited, as it requires storing all the possible program states associated with all possible execution paths. This will cause the path explosion problem~\cite{xie2009fitness}, especially for large functions with loops. Existing works try to address the pathfinding problem statically from an incomplete view, focusing on partial or local structures. For example, \textbf{DeepBinDiff}~\cite{duan2020deepbindiff} and \textbf{InnerEye}~\cite{zuo2018neural} match the CFGs based on semi-exhaustive path comparison, which is not scalable, and also misses the iterative graph learning.
\textbf{Genius}~\cite{xu2017neural}, \textbf{BinGo}~\cite{chandramohan2016bingo}, and \textbf{Tracelet}~\cite{david2014tracelet} use partial path matching, which lacks robustness when programs are easily altered through artificial means.
\textbf{BinaryAI}~\cite{yu2020order} uses graph convolution for message passing. However, this approach does not consider mutually exclusive dependencies among edges, covering invalid paths. The message passing mechanism also assumes a static adjacency matrix, which lacks high-level guidance from a global state.
The current research in this domain lacks a dedicated way to simulate the program state transitions along the guided valid execution path, with a focus on a higher order of node neighbourhood proximity. 

Inspired by symbolic execution for path-finding, we propose a neural network model, DeepEXE, which mimics a program state-guided execution process over the CFG to detect binary code vulnerabilities at the function or file level. 
DeepEXE relies on an execution agent that simulates and learns which direction to take, resulting in simulated paths across different epochs.
The combined node embedding represents the program state, and the branching actions guiding the program flow are based on the program state and code semantics of the current node.
DeepEXE leverages the implicit neural network paradigm, where only the final program state is stored before back-propagation. This enables a large simulation step over the execution flow.  
Compared to the existing methods with only local or partial graph information, DeepEXE enables modelling on the highest global-level view over the execution path.
Our contributions are as follows:

\squishlisttwo

\item We propose DeepEXE, a neural program execution model over a CFG for binary vulnerability detection. It simulates a semantic-guided decision process for stepping through a given function's CFG.

\item To simulate the program execution steps over the graph, we propose a learning agent for making branching decisions with an implicit neural network structure for program state transitions. It enables modelling program semantics on a higher level views over the execution path. 

\item To address the scalability and limited receptive field of graph neural networks, we use the implicit deep learning paradigm for nearly infinite message passing, which significantly enables global information aggregation in the graph and reduces the memory footprint.

\item We conduct experiments on two semi-synthetic datasets and two real world vulnerability datasets. We compare our methods against several state-of-the-art approaches and show that DeepEXE can consistently outperform the baselines in all scenarios.
\squishlistend

\section{Related Work}
\textbf{Vulnerability Detection}
While vulnerability detection can be conducted at either the source code or binary code level, we will discuss them together, since most methods can be applied to both levels, with some modifications. Machine learning-based (non-deep learning) methods involve the manual extraction of metrics and the input of these metrics as features~\cite{gupta2021extracting,sultana2021using}. The metrics can be multi-level and leverage the complexity characteristics of a program, such as the number of nested loops within a function. Manual feature extraction is more expensive and requires expert knowledge. Also, the features need to be constantly updated to accommodate changes in the codebase. Text-based deep learning is very popular for source code vulnerability detection, where different granularity levels can be leveraged in order to obtain text features or embeddings. Li et al. group tokens based on semantics and syntax into slices or gadgets~\cite{li2021vuldeelocator, li2018vuldeepecker, zou2019mu}, and feed them into a LSTM model. For binary code, Instruction2Vec~\cite{b8} and Bin2img~\cite{b10} use instruction embedding as a preprocessing step. Similar to Word2Vec, the embedding contains contextual dependency and can be used to detect vulnerabilities at a later stage, which is a 1D CNN model. These models solely focus on the semantics of the tokens, where the structural information is omitted. There are several GNN models at the source code that use different graphs that can be parsed from source code, such as abstract syntax trees, data dependence graphs, and control flow graphs~\cite{csahin2022predicting,zhou2019devign, cao2021bgnn4vd}. For GNN message passing, there are multiple styles that we will discuss next.

\textbf{Graph Neural Networks and Implicit Models}
In binary code, GNN methods aim at learning the structures by first parsing the assembly code into control flow graphs and performing message passing. There are multiple variants related to graph neural networks. The pioneer works of graph neural networks are mostly associated with recurrent graph neural networks ~\cite{gori2005new, scarselli2008graph, gallicchio2010graph, li2015gated, dai2018learning}, where the node representations are aggregated with a fixed set of parameters. Convolutional graph neural networks~\cite{kipf2016semi, hamilton2017inductive, velivckovic2017graph} expand the GNN by using multiple layers with different parameters. This approach addresses the cyclic mutual dependencies architecturally~\cite{wu2020comprehensive} and is more efficient and powerful. However, GNNs struggle to capture long-range dependencies in large graphs, due to the finite number of message passing iterations. One potential solution is the recently studied implicit neural networks. The implicit learning paradigm is different from traditional deep learning, as it solves the solution for a given equilibrium problem, which is formulated as an nearly infinite layer network. Implicit models have previously shown success in domains such as sequence learning~\cite{bai2019deep}, physics engine~\cite{de2018end}, and graph neural networks ~\cite{gu2020implicit}.


\section{Preliminaries}
In this section, we define some necessary notation involving our learning problem, including the input and output. We provide further discussion about the graph neural network and the reinforcement learning in Appendix~\ref{appendix: preliminaries}.


\textbf{CFGs and Basic Blocks} The input of the model is a binary file in assembly code. The assembly functions and their CFGs are both obtained from the IDA Pro disassembler~\footnote{\href{https://hex-rays.com/ida-pro/}{IDA Pro}}. Each function is regarded as a graph $\gG$ that contains segmented code blocks called basic blocks, which are sequences of instructions without any jump or call to other blocks. As the input to the neural network, a graph $\gG = (\mV, \mA)$ has the blocks $\mV \in \mathbb{R}^{n \times v}$ with $n$ nodes, $v$ tokens, and the adjacency matrix $\mA \in \mathbb{R}^{n \times n}$. $\mA$ defines all directed edges within the graph and is obtained by extracting call statements between the blocks. Note that $\mA$ has 0 across the diagonal element and is non-symmetrical. Moreover, we apply the re-normalization trick to $\mA$~\cite{kipf2016semi}, in order to prevent numerical instabilities during deep network training. For file level classification, we merge the function graphs as a whole, based on the function call information. Moreover, additional information, such as comments and names, are removed. The basic block $\mV$ only contains operations and operands of instructions. 


\textbf{Problem Statement} We define several neural network modules within our architecture ${F} = ({F_S}, F_I, {F_A})$, where $F_S$ is the sequential model for semantics embedding, $F_I$ is the implicit graph neural network model for structure and node embedding, and $F_A$ is the reinforcement learning agent for dynamic pathing optimizer, given certain program states. The goal is to predict whether each function contains a vulnerability. Given the input graph $\gG = (\mV, \mA)$, our model learns several levels of information and aggregates them together for the final output of the model, which is a binary classification score $F: \gG \rightarrow \hat{y} \in \mathbb{R}$. Formally, we define the following learning task parameterized by $\theta$:

\begin{equation}
    \hat{y} = \operatorname*{argmax}_{y' \in {0,1}} F_\theta(y'|\gG, \theta);    
    \theta = \operatorname*{argmax}_\theta F_\theta(y'=y|\gG, \theta)
\end{equation}

\begin{figure*}[t]
\centering
  \includegraphics[width=\textwidth]{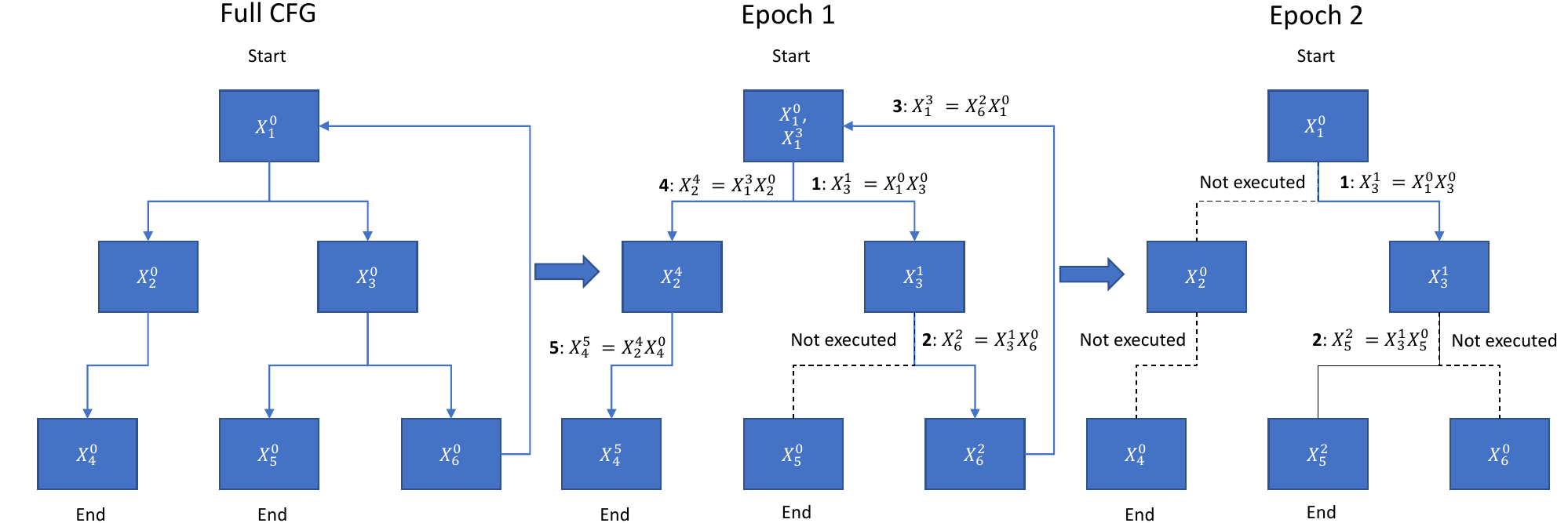}
  \caption{In each epoch, the model simulates one execution session with a specific execution path consisting of multiple steps. At step $i$, the executor chooses the most likely branch for node $j$ to move next based on program state and node semantics. This is one execution with a loop (Epoch 1) and one without (Epoch 2). The model then updates the program state by combining the next node's code semantics.
  }
  \label{fig:model}
\end{figure*}

\section{Neural Control Flow Execution}
We design the DeepEXE architecture with semantic-driven and execution-guided principles. 
CFGs extracted from disassembly contain crucial information about the program logic and paths, which dictates the outputs and functionalities of assembly code. An important characteristic to differentiate CFGs from graphs in other domains, such as social networks or chemistry, is that node states should be dependent on the execution logic. Programs are executed following specific orders based on the dependencies among the edges conditioned by the program state, where the results and semantics can substantially differ when orders vary. 


We borrow the idea of symbolic execution~\cite{baldoni2018survey} and create a neural CFG executor.
A training epoch contains a full iteration of the executive session, which corresponds to a concrete execution path. 
Note that each epoch can have completely different execution paths, as the model learns. The overall architecture is shown in Appendix~\ref{appendix:arch}.

An example of the learning process is shown in Figure~\ref{fig:model}. 
The training consists of many training epochs. In Figure~\ref{fig:model}, the path for Epoch 1 goes into a loop, while Epoch 2 directly goes into the exit point. 
The execution agent performs multiple steps within an epoch. It starts from the entry node, then transitions to other possible nodes in each step. 
The decision on which branch to select depends on the program state $X$, and 
$X^i_j$ indicates the updated program state at step $i$ for node $j$. After jumping to the next node, the agent updates the program state and repeats the decision process until it reaches an equilibrium state. 


\subsection{Token Semantics}
In this section, we discuss the preprocessing, embedding, and sequential learning task. A basic block contains a stream of instructions, which can be further broken down into operations and operands, and be tokenized. We treat the entire block as one sentence and apply a subword and unigram~\cite{kudo2018subword,kudo2018sentencepiece} model for the token encoding, which mitigates the out-of-vocabulary problem. 
Assembly code is compiler dependent and can easily result in out-of-vocabulary (OOV) tokens. A way to address the OOV issue is to break down the tokens into characters for encoding. Even with a fixed vocabulary size, unseen tokens can be encoded by matching the subword to their closest known tokens. Moreover, it is not language dependent and can be trained from scratch very efficiently. 
We increase the subspace representation power by simply applying an embedding layer $E: \mV \rightarrow \mathbb{R}^{n\times v \times h}$ next, where $h$ is the hidden dimension. Note that we use $h$ as the hidden dimension throughout the paper for simplicity, but different dimensions can be used for any layers in practice. The sequential model used in this task is a bi-directional GRU~\cite{chung2014empirical}. The output of the GRU layer $\mU \in \mathbb{R}^{n \times v \times h}$ further embeds the token semantics by taking contextual information into account. In order to obtain a representation for the entire basic block, a maximum or average pooling along the time dimension is used to compute $\mU \in \mathbb{R}^{n \times h}$ for block embedding.

\subsection{Program State Guided Execution and Functional Representation}
\textbf{Program State} Initial node representation $U$ establishes the semantics within basic blocks, but it is not sufficient to simply globally aggregate $U$ for a high-level representation of the graph. In this regard, a reinforcement agent $a^{t}(s^{t-1})$ that decides the next execution path is defined, given the previous program state $s^{t-1}$. Unlike traditional neural networks that perform forward and backward pass one at a time, our approach internally loops through multiple states $t$ within a training epoch. We define the program state as a linear transformation of the node state $X^{t}$, where $X^{0} = U$, and some trainable parameter $W_{s} \in \mathbb{R}^{h \times 1}$:
\begin{equation}
    s^{t} = \sigma(X^tW_s)
\end{equation}
\textbf{Agent Reparameterization} Due to the backpropagation algorithm, categorical variables are hard to train in this stochastic environment in the neural network. This layer effectively becomes non-differentiable when using normal sampling process such as $\operatorname*{argmax}$. A solution is to use the Gumbel softmax~\cite{jang2016categorical} to re-parameterize the state while maintaining the ability to backpropagate efficiently during training. Gumbel softmax is a continuous and differentiable distribution that can sample categorical distribution, it is given by:
\begin{equation}
    z_i^t = \frac{exp((log(s_i^{t-1}) + g_i)/\tau )}{\Sigma^k_jexp((log(s_j^{t-1}) + g_j)/\tau}, \textbf{for}~i = 1,...,k
\end{equation}
where $z_i^t$ is the sample drawn from the state, $g_i \sim Gumbel(0,1)$ are samples drawn i.i.d from the Gumbel distribution, and $\tau$ is the temperature controlling the discreteness of the new samples. Gumbel softmax works better with a lower value for $\tau \in [0, \infty]$ as it approaches to $\operatorname*{argmax}$ smoothly, whereas setting a large value makes the samples become uniform.

\textbf{Adjacency Matrix Update} In each state update, the agent walks through the graph with updated program state to capture the intermediate execution path that leads to certain results. 
We have the flexibility to design the agent to be either hard or soft. A soft agent $a^t = z^t$ preserves the probabilities drawn from Gumbel softmax, which implies that a program information can flow in different execution paths at the same time based on the probabilities $\sum_{i} z^t_i = 1$. A hard agent mimics the execution path and is one-hot, leading to one strictly one execution at a time. 
The agent $a^{t} \in \mathbb{R}^{n \times 1}$ is then used to select a path and generate the state-dependent adjacency matrix $\Tilde{\mA}^{t}$, which is updated as: $\Tilde{\mA}^{t} = \mA a^{t}$.


\subsection{Executor Stepping Via Implicit GNN}
\label{sec:step}
\textbf{Implicit GNN} With the updated adjacency matrix from the agent, one can perform graph neural network on the CFG to aggregate neighbour information into the nodes. However, assembly code can be large for various reasons. For example, a GCC compiler can use an optimization level that minimizes the execution size and reduces the size of CFGs. While GNN is a suitable approach to learn the structural dependency of a function, it requires a pre-defined number of layers, where each layer usually performs 1-hop message passing. Intuitively, the vanilla GNNs do not scale well with large graphs and can fail to capture global information. The dependency between further nodes can be crucial to understand the overall semantics of a program. Such long range dependency is difficult to capture with longer edges.
To alleviate the above stated problem, we perform the program state transitions in an implicitly defined style. In general, the transition at state $t$ can be written as an implicit form of the GNN layer:
\begin{equation}
\label{eqn:general}
    X^{t+1} = \phi(X^{t}W^t\mA^{t}+U) 
\end{equation}
\begin{equation}
\label{eqn:prediction}
    y'=f_\psi(X^*)
\end{equation}
Such form of layer does not explicitly output a vector to be fed into the next layer. Instead, it uses a fixed point iteration in equation~\ref{eqn:general} that aims to find the equilibrium vector state $X^*$ as $t \rightarrow \infty$. The equilibrium state is then used for the prediction task in equation~\ref{eqn:prediction}, where $f_\psi$ is an output function parameterized by $\psi$ for the desired classification task.
With the reinforcement agent embedded in the updated adjacency matrix $\Tilde{\mA}^*=\Tilde{\mA}^t:t\rightarrow \infty$, our equilibrium solution is formulated as follows:
\begin{equation}
\label{eqn:implicit}
    X^* = \phi(X^*W\Tilde{\mA}^*+b_{\Omega}(U))
\end{equation}
$W\in\mathbb{R}^{h\times h}$ and $\Omega\in\mathbb{R}^{h\times h}$ are parameters, and $U$ is the initial node feature. 
Note that only a single layer is required to produce the updated node representation $X$ iteratively instead of multiple stacking layers. We also inject $U$ into the equation through some affine transformation $b_{\Omega}$. This ensures that original node semantics is preserved throughout the iterations when solving for the fixed point~\cite{bai2019deep}.

\textbf{Fixed Point Acceleration} Although the equilibrium point can be obtained from iterating equation~\ref{eqn:implicit} infinitely, it is not the most efficient and stable method for convergence. More importantly, it does not guarantee convergence. Anderson acceleration~\cite{walker2011anderson} is an accelerated algorithm for finding fixed points. Given a function $f$ to solve, which is equation~\ref{eqn:implicit} in our case, we define (1) $m_k = min\{m, t\}$ as the parameter for controlling past iteration memory by setting $m$ to any positive integer; (2) $g(x) = f(x) - x$ as the residual with the matrix $G_t = [g_{t-m_t}, ..., g_{t}]$. The root solving process using Anderson acceleration is formulated as:
\begin{equation}
\begin{split}
 \begin{gathered}
    \alpha_t = \operatorname*{argmin}_{\alpha}||G_t\alpha||_2,  \\
    \textbf{where}~\alpha=(\alpha_0,...,\alpha_{m_t}) \in \mathbb{R}^{m_t+1}:\sum_{i=0}^{m_t}\alpha_i=1
 \end{gathered}
\end{split}
\end{equation}
\begin{equation}
    x^{t+1} = \sum_{i=0}^{m_t}(\alpha_t)_if(x_{t-m_t+i})
\end{equation}
Instead of computing for $x^{t+1}$ directly from $x^t$, Anderson acceleration solves for a coefficient $\alpha$ in an optimization problem that minimizes the norm of $g(x)$.


\textbf{State Transition Termination} The executor terminates in three different scenarios. (1) If the executor reaches the exit point on the CFG, there will not be any updates to $X^{t+1}$ after Equation~\ref{eqn:implicit}, naturally leading to an equilibrium state. (2) If the executor reaches an equilibrium state, but not at the program exit point, it logically indicates that further execution will not result in changes in the program state. Therefore, it is natural to terminate. (3) If the executor reaches a configured maximum steps.

Once $X^*$ is at equilibrium, we apply layer normalization~\cite{ba2016layer} and global average pooling layer to obtain the graph representation $\mG$:
\begin{equation}
    \mG = \textbf{LayerNorm}(\frac{\sum_i^nX^T_{i,j}}{n}), \forall j = 1,...,h
\end{equation}
The prediction task can be simply computed by a linear transformation to get the logits:
\begin{equation}
    y' = W_p\mG, \textbf{where}~W_p \in \mathbb{R}^{1 \times h}
\end{equation}
We want to emphasize that through the use of an implicitly defined GNN layer, it is no longer required to have multiple stacking GNN layers to achieve higher order node aggregation. Instead, each state transition within the layer effectively performs a message passing, as a normal GNN layer would. This has the benefits of lowering the memory costs, while maintaining the same level of representational power, given similar parameter count. Moreover, the long range dependency issue can be effectively addressed by iterating a nearly infinite number of state transitions. 

\subsection{Training}
While the forward pass in an implicit network possesses some nice properties for the network discussed earlier, it is not a trivial task to train the backward pass. Traditionally, a neural network contains exact operations with explicitly defined input and output, where the gradients can be computed via chain rule. We first define the loss term $l$:
\begin{equation}
    l = \mathcal{L}(\hat{y},y) = \mathcal{L}(F_{\psi}(\mG), y)
\end{equation}
$F_{\psi}$ is the prediction rule that takes the graph embedding $\mG$. $\mathcal{L}(\cdot)$ computes the cross entropy loss and outputs the scalar $l$. Using chain rule, the loss can be backpropagated as:
\begin{equation}
    \frac{\partial l}{\partial \theta} = \frac{\partial l}{\partial \mG} \frac{\partial \mG}{\partial X^*} \frac{\partial X^*}{\partial \theta}
\end{equation}
The terms $\frac{\partial l}{\partial \mG}$ and $\frac{\partial \mG}{\partial X^*}$ can be both computed using any autograd software. However, the term $\frac{\partial X^*}{\partial \theta}$ is difficult to compute, since the equilibrium point $X^*$ is obtained through iterative root finding. If we unroll this computation graph, the network needs to store all intermediate gradients for every state transition. Depending on the number of transitions, it is not a practical approach. Instead, we write $X^*$ in its implicitly defined form:
\begin{equation}
    X^*(\theta) = \phi(X^*W\Tilde{\mA}^*+b_{\Omega}(U)) = F_{I}(X^*(\theta), U)
\end{equation}
where $F_{I}$ denotes the implicit graph neural network. By taking the derivative with respect to $\theta$, we obtain:
\begin{equation}
\label{eqn:train1}
    \frac{\partial X^*(\theta)}{\partial \theta} = \frac{\partial F_{I}(X^*(\theta), U)}{\partial \theta}
\end{equation}
By applying the chain rule on the right hand side of equation~\ref{eqn:train1}, we expand it into the following:
\begin{equation}
    \frac{\partial X^*(\theta)}{\partial \theta} = \frac{\partial F_{I}(X^*, U)}{\partial \theta} + \frac{\partial F_{I}(X^*, U)}{\partial X^*}\frac{\partial X^*(\theta)}{\partial \theta}
\end{equation}
At this point, both $\frac{\partial F_{I}(X^*, U)}{\partial \theta}$ and $\frac{\partial F_{I}(X^*, U)}{\partial X^*}$ can again be obtained using autograd software. The last unknown term $\frac{\partial X^*(\theta)}{\partial \theta}$ is computed by solving the linear system. In our approach, we use Anderson acceleration to iteratively solve this term. 

Through implicit differentiation, we directly evaluate the gradient at the equilibrium point. We avoid the computation of any intermediate state transition and can efficiently backpropagate through the network, even with a nearly infinite number of transitions. This also has a better memory footprint.


\begin{table*}[t]
\centering
\caption{NDSS18 Dataset Evaluation}
\label{tab:ndss}
\begin{tabularx}{1\textwidth}{c @{\extracolsep{\fill}} cccccc} 
 Models & Input Type & Accuracy & Recall & Precision & F1 & AUC \\
 \hline
 \hline
    Bi-LSTM & Assembly Ins. & 85.38 & 83.47 & 87.09 & 85.24 & 94.89 \\ 
    GCN & CFG & 86.48 & 84.59 & 88.12 & 86.32 & 95.81 \\
    MD-CWS~\cite{le2018maximal} & Assembly Ins. & 85.30 & 98.10 & 78.40 & 87.10 & 85.20 \\
    MD-CKL~\cite{le2018maximal} & Assembly Ins. & 82.30 & 98.00 & 74.80 & 84.00 & 82.10 \\ 
    MD-RWS~\cite{le2018maximal} & Assembly Ins. & 83.7 & 94.3 & 78.0 & 85.4 & 83.5 \\
    MDSAE-NR~\cite{albahar2020modified} & Assembly Ins. &  87.50 & \textbf{99.30} & 81.20 & 89.80 & 87.10 \\
    TDNN-NR~\cite{albahar2020modified} & Assembly Ins. & 86.60 & 98.70 & 80.30 & 88.30 & 86.30 \\
    VulDeePecker~\cite{li2018vuldeepecker} & Source Code Gadgets & 83.50 & 91.00 & 79.50 & 84.80 & 83.40 \\
    DeepEXE & CFG & \textbf{90.58} & 89.36 & \textbf{92.13} & \textbf{90.72} & \textbf{98.01} \\
    \hline
\end{tabularx}
\end{table*}

\begin{table*}[t]
\centering
\caption{Juliet Dataset Evaluation}
\label{tab:juliet}
\begin{tabularx}{.8\textwidth}{c @{\extracolsep{\fill}} cccccc} 
 Models & Input Type & Accuracy & Recall & Precision & F1 & AUC  \\
 \hline
 \hline
    Bi-LSTM & Assembly Ins. & 96.81 & 98.44 & 95.48 & 96.94 & 99.03 \\ 
    gcn~\cite{b9} & CFG & 97 & NA  & NA  & NA  & NA  \\ 
    i2v/CNN~\cite{b8} & Assembly Ins. & 87.6 &N/A &N/A &N/A &N/A \\ 
    i2v/TCNN~\cite{b8} & Assembly Ins. & 96.1 &N/A &N/A &N/A &N/A \\ 
    w2v/CNN~\cite{b8} & Assembly Ins. & 87.9 &N/A &N/A &N/A &N/A \\ 
    w2v/TCNN~\cite{b8} & Assembly Ins. & 94.2 &N/A &N/A &N/A &N/A \\ 
    i2v~\cite{b10} & Assembly Ins. & 96.81 & 97.07 & 96.65 & 96.85 &N/A \\ 
    bin2img~\cite{b10} & Assembly Ins. & 97.53 & 97.05 & 97.91 & 97.47 &N/A \\ 
    w2v~\cite{b10} & Assembly Ins. & 96.01 & 96.07 & 95.92 & 95.99 &N/A \\ 
    DeepEXE & CFG & \textbf{99.80} & \textbf{99.60}  & \textbf{100.00}  & \textbf{99.80}  & \textbf{100.00}  \\ 
    \hline
\end{tabularx}
\end{table*}

\section{Experiment}
In this section, we demonstrate the ability of DeepEXE on predicting binary code vulnerability in a variety of scenarios. To properly evaluate DeepEXE, we conduct experiments using two semi-synthetic datasets and two real world datasets. The NDSS18\footnote{https://samate.nist.gov/SRD/index.php, Software Assurance Reference Dataset} and Juliet Test Suites\footnote{https://samate.nist.gov/SARD/test-suites, NIST Test Suites} are both semi-synthetic datasets commonly used as for vulnerability detection tasks. Though the practical implications for a method should not solely depend on the synthetic results, as they are less complex. For real world datasets that are larger and can contain less trivial vulnerabilities, we employ the FFmpeg\footnote{https://ffmpeg.org/, FFmpeg} and Esh~\cite{david2016statistical} datasets. The details of the datasets can be found in Appendix \ref{appendix:dataset}. For the baseline methods, we inherit the results reported in previous works, due to the large amount of experiments and different setups. The evaluation metrics reported include accuracy, precision, recall, F1 score, and area under the ROC curve (AUC). We randomly split each dataset into 75\% for training and 25\% for evaluation. Some metrics are not shown in the baselines because of their absence in the original works. 


\subsection{Evaluation}
For each dataset, we compare DeepEXE with the benchmarks that are also evaluated on the same dataset due to limited space. The details of each baseline is described in Appendix \ref{appendix:baseline}. Additionally, we build baseline models that work inherently well in this task including bi-directional LSTM (Bi-LSTM)~\cite{hochreiter1997long} and graph convolution network (GCN)~\cite{kipf2016semi}.

\textbf{Semi-Synthetic Results} We first analyze the results for the NDSS18 dataset shown in Table.~\ref{tab:ndss}. The two baselines we implemented (Bi-LSTM and GCN) have surprisingly comparable results with the benchmarks, including MDSAE-NR and TDNN-NR. All the MDSAE-based methods have imbalanced precision and recall, where the models tend to overestimate the vulnerable code. DeepEXE has the best overall performance, leading the accuracy and AUC by ~3\%. Moreover, DeepEXE is a CFG-based method and we empirically show that by adding the execution-guided agent and expanding the receptive field of graph convolution, it is able to capture more topological information. Note that even in a scenario where the recall metric is highly important, the classification threshold can always be adjusted to accommodate the balance between precision and recall. We are also able to outperform VulDeePecker, which is a source code level method that only leverages the sequential information of the code gadget, potentially omitting the much useful topological knowledge of the source code.

\begin{table*}[t]
\centering
\caption{FFmpeg Dataset Evaluation}
\label{tab:ffmpeg}
\begin{tabularx}{1\textwidth}{c @{\extracolsep{\fill}} cccc} 
 Models & Code Level & Input Type & Accuracy & F1  \\
 \hline
 \hline
    Bi-LSTM~\cite{zhou2019devign} & Source Code & Code Snippets & 53.27 & 69.51  \\ 
    Bi-LSTM + Attention~\cite{zhou2019devign} & Source Code & Code Snippets & 61.71 & 66.01 \\
    CNN~\cite{zhou2019devign} & Source Code & Code Snippets & 53.42 & 66.58  \\
    GGRN-CFG~\cite{zhou2019devign} & Source Code & CFG & 65.00 & 71.79 \\ 
    GGRN-composite~\cite{zhou2019devign} & Source Code & AST, CFG, DFP, NCS & 64.46 & 70.33 \\
    Devign-CFG~\cite{zhou2019devign} & Source Code & CFG &  66.89 & 70.22 \\
    Devign-composite~\cite{zhou2019devign} & Source Code & AST, CFG, DFP, NCS & \textbf{69.58} & \textbf{73.55} \\
    DeepEXE & Binary Code & CFG & 68.29 & 67.17 \\
    \hline
\end{tabularx}
\end{table*}

\begin{table*}[t]
\centering
\caption{Esh Dataset Evaluation}
\label{tab:esh}
\begin{tabularx}{.8\textwidth}{c @{\extracolsep{\fill}} cccccc} 
 Models & Input Type & Accuracy & Recall & Precision & F1 & AUC  \\
 \hline
 \hline
    Bi-LSTM & Assembly Ins. & 99.49 & 79.48 & 88.57 & 83.78 & 96.87 \\ 
    GCN & CFG & 99.31 & 63.89 & \textbf{95.83} & 76.67 & 83.54 \\
    DeepEXE & CFG & \textbf{99.78} & \textbf{95.65} & 91.67 & \textbf{93.62} & \textbf{99.78} \\
    \hline
\end{tabularx}
\end{table*}

The Juliet dataset evaluation is shown in Table~\ref{tab:juliet}. As a synthetic dataset, the test cases contain much shorter code. However, there are over 100 different CWEs among all test cases. In reality, a detection tool should be robust enough to detect unseen or zero-day vulnerabilities. It is useful for evaluating the robustness and generalizability of an approach. DeepEXE shows nearly perfect detection accuracy and AUC for this dataset. This shows that even with the single-layer design, DeepEXE is able to generalize well enough. As the graphs are usually small in these test cases, the execution paths generated by static analysis are likely more accurate. Therefore, we believe the implicit GNN contributes more to the performance increase than the agent in this case.

\textbf{Real CVE Results} We evaluate the FFmpeg dataset shown in Table~\ref{tab:ffmpeg}, which specifies the code levels and input types. Since Devign detects vulnerabilities at the source code level, it is significantly easier with the rich semantics, syntax, and structures. DeepEXE is able to outperform most of the approaches, even at the binary code level. In particular, when only using the CFG as input, DeepEXE achieves better accuracy than both the Devign and GGRN models. Devign-composite utilizes multiple input graphs, such as AST, DFP and NCS. These additional graphs are usually only available for source code. DeepEXE shows its capability at detecting vulnerabilities for real-world and complex programs. Moreover, source code CFGs are less complicated to generate, whereas binary CFGs often can be an over-estimation of the true control flow. With our execution-guided approach, we limit the errors caused by such approximation, while maintaining a high level of global information. The receptive field of GNN in DeepEXE is practically unlimited, allowing us to accommodate for much larger graphs. 

Lastly, we show the evaluation results for the Esh dataset in Table~\ref{tab:esh}. Due to the extreme imbalance of labels distribution, which is the case in many real-life scenarios, the Bi-LSTM and GCN baselines have lower recalls. The recall metric is important when there are fewer vulnerable cases. DeepEXE, on the other hand, is able to distinguish vulnerable code from non-vulnerable code, given the small number of positive labels. Note that the class weight is not manually adjusted during training, as it is cumbersome and inefficient to tune it for every dataset in practice. With over 90\% percision, DeepEXE is able to identify 95\% of the vulnerable CVE cases. Similar to FFmpeg, although many cases in the Esh dataset contain a large number of nodes, DeepEXE is inherently designed to handle such large graphs and outperform other baselines.

\section{Conclusions}
We have proposed DeepEXE, a control flow execution-guided deep learning framework for binary code vulnerability detection. Given the importance of binary code learning, we address two major gaps in the existing research works, which are the lack of modelling program state transition and scalability for large graphs. Instead of assuming the CFG is accurate, which is often not the case, due to the over-estimation from static analysis, we use a reinforcement agent to guide the execution of a program flow that mimics the behaviour of dynamic analysis. DeepEXE is able to capture certain program state transitions that lead to specific vulnerability results, creating a higher dependency between the output and internal node state and topological information. 
We also show the benefits of training an implicitly defined network, which are directly obtaining the gradients for the equilibrium point and mitigating the heavy memory footprint in large networks. In the experiments, we demonstrate that DeepEXE outperforms all state-of-the-art vulnerability detection methods for the NDSS18 and Juliet datasets. DeepEXE is also very competitive in detecting real world CVEs, even when compared to source code level methods, which are less difficult, given the amount of available information. Overall, DeepEXE is a robust and accurate tool for binary vulnerability detection.

In the future, there are several potential directions to grow for DeepEXE. First of all, the training time is slower than for the traditional neural network, due to the many iterations for obtaining equilibrium. This can be improved by using more sophisticated solvers to reduce the number of steps for equilibrium computation. Next, DeepEXE does not have to be restricted to vulnerability detection in the cybersecurity domain. For other security tasks, such as binary code similarity comparison or malware detection, matching the graph structures of malicious programs is often done using GNN. By modifying the training objective, DeepEXE can be used for a lot more of supervised and unsupervised tasks. Moreover, as long as the input data has some form of graphical structures, we can apply the same design to many other domains, such as social network and chemistry studies.

\bibliographystyle{ACM-Reference-Format}
\bibliography{ref}

\appendix
\section{Appendix}

\subsection{Architecture}
\label{appendix:arch}
\begin{figure*}[t]
\centering
  \includegraphics[width=\textwidth]{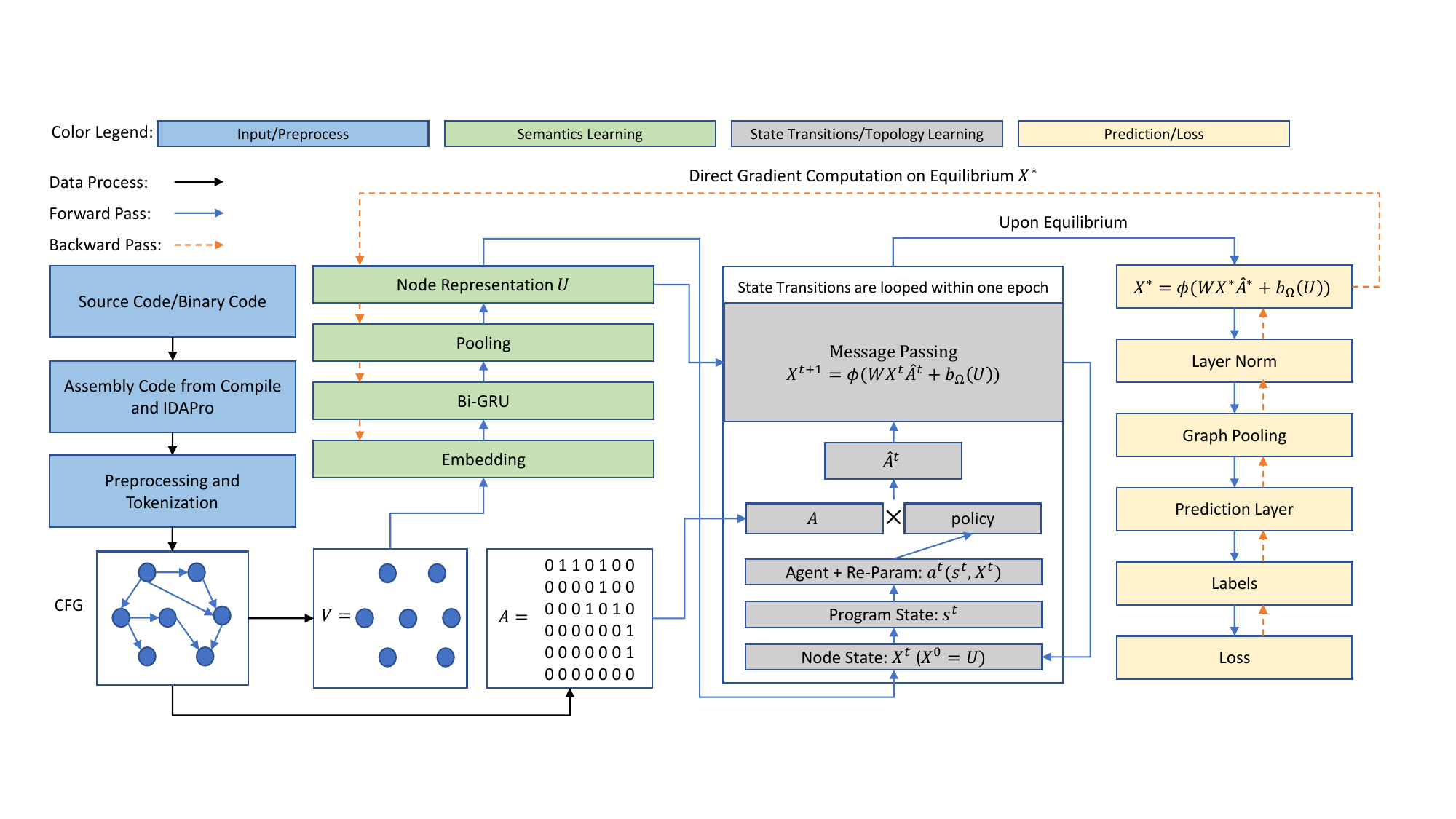}
  \caption{Overall architecture of DeepEXE. Four major segments of our model include input preprocessing, node embedding through sequential model, state transition and structure learning, and prediction and training. DeepEXE combines the local instruction semantics with high-level topological information, where program dependencies are captured through the use of a REINFORCE agent and a GNN with much larger receptive field.
  }
  \label{fig:architecture}
\end{figure*}

We show the overall architecture including the input preprocessing, semantics learning, state transition, and predictions and training in Figure~\ref{fig:architecture}.

\subsection{Datasets and Setup}
\label{appendix:dataset}

The hardware used for the experiments includes a RTX6000 GPU, Intel Xeon Gold 5218 CPU, and 64GB of memory. The main software used includes Python 3.9.10 and PyTorch 1.10.2 on Ubuntu 20.04.3 LTS.

\textbf{Semi-Synthetic Datasets} include the NDSS18 dataset and Juliet Test Suite. The NDSS18 dataset is a derivation from the National Institute of Standards and Technology (NIST): NVD\footnote{https://nvd.nist.gov/, National Institute of Standards and Technology} and the Software Assurance Reference Dataset (SARD) project\footnote{https://samate.nist.gov/SRD/index.php, Software Assurance Reference Dataset}. NDSS18 was first published by~\cite{li2018vuldeepecker} as a source code vulnerability dataset and later compiled to binary code by~\cite{le2018maximal} for binary level detection. It includes a total of 32,281 binary functions that are compiled using Windows and Linux. There are two types of Common Weakness Enumerations (CWEs)\footnote{https://cwe.mitre.org/, Common Weakness Enumeration (CWE)} in NDSS18: CWE119 and CWE399. Juliet Test Suite is a collection of 81,000 test cases in C/C++ and Java from NIST\footnote{https://samate.nist.gov/SARD/test-suites, NIST Test Suites} that contain 112 different CWEs. Both datasets have nearly balanced distributions for the labels.

\textbf{Real CVE Datasets} include the FFmpeg vulnerabilities and Esh datasets, which are both extracted from real world applications or open-source libraries. The codebase is significantly larger than the ones in semi-synthetic datasets. Vulnerabilities are often harder to detect in these programs, due to the much increased complexity. FFmpeg\footnote{https://ffmpeg.org/, FFmpeg} is an open-source suite of libraries written in C for handling media files, such as video and audio. It was first used in source code vulnerability detection~\cite{zhou2019devign}, where the authors manually collected and labelled the data for various vulnerability commits on Github. We compile the FFmpeg source code provided by the authors into binary code and obtain 16,494 binary functions, where 7,257 are vulnerable and 9,237 are non-vulnerable. The Esh dataset contains CVE cases collected by David et al.~\cite{david2016statistical}, which include 8 different CVEs: cve-2014-0160, cve-2014-6271, cve-2015-3456, cve-2014-9295, cve-2014-7169, cve-2011-0444, cve-2014-4877, and cve-2015-6826. In total, there are 3,379 cases and only 60 are vulnerable. The distribution of vulnerability in the Esh dataset is highly imbalanced, which represents a more realistic scenario.

\subsection{Baselines and hyperparameters}
\label{appendix:baseline}
\textbf{NDSS18 baselines} Maximal Divergence sequential Autoencoder (MDSAE) was proposed by~\cite{le2018maximal} and uses a deep representation learning approach. The model aims to discriminate the vulnerable and non-vulnerable code by forcing it to be maximally divergent. The input to MDSAE is a sequence of binary code instructions. We report the three best variants in this paper. MMDSAE~\cite{albahar2020modified} is a modified version inspired by MDSAE and uses a similar approach that adds a regularization technique. The authors propose two variants, namely MDSAE-NR and TDNN-NR, that have similar performance. The last baseline we include for this dataset is VulDeePecker~\cite{li2018vuldeepecker}, which is a source code vulnerability detection. Instead of using binary code instructions, VulDeePecker takes source code gadgets as input, which are blocks of code with tightly associated semantics. Although this is a source code evaluation, the underlying dataset used is the same. We therefore include VulDeePecker as one of the baselines. It also only uses sequences of code as input and does not consider any topological information from the code graphs.

\textbf{Juliet baselines} Bin2Vec~\cite{b9} is a graph-based binary vulnerability detection approach that utilizes graph convolution network by taking the CFG as input. Instruction2Vec~\cite{b8} is a representation learning approach to embed the assembly instructions into vectors and apply the downstream vulnerability detection task. The instruction embedding is similar to Word2Vec, it utilizes different parts of an instruction and combines them as a single vector. The downstream vulnerability detection is achieved by training a CNN or Text-CNN using the vectors. The same authors later proposed an updated version of Instruction2Vec~\cite{b10} and includes a few variants, including Word2Vec and Binary2Img. All Instruction2Vec related methods use the assembly instructions as input and do not consider the structural information.

\textbf{FFmpeg baselines} For the FFmpeg dataset, we compared DeepEXE to Devign~\cite{zhou2019devign}, which provides the source code for FFmpeg. Devign uses a gated graph recurrent network (GGRN)~\cite{ruiz2020gated} as a graph learning technique. Unlike binary code, where only CFG can be extracted, Devign detects vulnearbilities at the source code level. It takes several intermediate graph representations of source code, such as AST, CFG, DFG, and NCS. We include several variants of Devign, such as Bi-LSTM, GGRN with CFG or all graphs, and Devign with CFG or all graphs. Note that we directly compare our results on the binary code with the original results of Devign, which are based on source code. 

\textbf{Esh baselines} To the best of our knowledge, Esh is not used in any other papers for vulnerability detection evaluation. The original paper that provided this dataset evaluates it at the basic block level and focuses on code matching. Therefore, we compare DeepEXE to the Bi-LSTM and GCN baselines we implemented ourselves.

For the baseline methods, we directly inherit the results reported in previous works, due to the large amount of experiments and different setups. The evaluation metrics reported include accuracy, precision, recall, F1 score, and area under the ROC curve (AUC). Cross-validation is used for tuning hyperparameters in order to obtain optimal accuracy and reasonable memory usage. We use a universal hidden dimension of 64, learning rate of 0.01 with the Adam optimizer~\cite{kingma2014adam}, dropout rate of 0.5, batch size of 192, and a maximum iteration of 50 for the Anderson acceleration solver. We randomly split each dataset into 75\% for training and 25\% for evaluation. Some metrics are not shown in the baselines because of their absence in the original works. The hardware used for the experiments includes a RTX6000 GPU, Intel Xeon Gold 5218 CPU, and 64GB of memory. The main software used includes Python 3.9.10 and PyTorch 1.10.2 on Ubuntu 20.04.3 LTS.

\textbf{NDSS18 baselines} Maximal Divergence sequential Autoencoder (MDSAE) was proposed by~\cite{le2018maximal} and uses a deep representation learning approach. The model aims to discriminate the vulnerable and non-vulnerable code by forcing it to be maximally divergent. The input to MDSAE is a sequence of binary code instructions. We report the three best variants in this paper. MMDSAE~\cite{albahar2020modified} is a modified version inspired by MDSAE and uses a similar approach that adds a regularization technique. The authors propose two variants, namely MDSAE-NR and TDNN-NR, that have similar performance. The last baseline we include for this dataset is VulDeePecker~\cite{li2018vuldeepecker}, which is a source code vulnerability detection. Instead of using binary code instructions, VulDeePecker takes source code gadgets as input, which are blocks of code with tightly associated semantics. Although this is a source code evaluation, the underlying dataset used is the same. We therefore include VulDeePecker as one of the baselines. It also only uses sequences of code as input and does not consider any topological information from the code graphs.

\textbf{Juliet baselines} Bin2Vec~\cite{b9} is a graph-based binary vulnerability detection approach that utilizes graph convolution network by taking the CFG as input. Instruction2Vec~\cite{b8} is a representation learning approach to embed the assembly instructions into vectors and apply the downstream vulnerability detection task. The instruction embedding is similar to Word2Vec, it utilizes different parts of an instruction and combines them as a single vector. The downstream vulnerability detection is achieved by training a CNN or Text-CNN using the vectors. The same authors later proposed an updated version of Instruction2Vec~\cite{b10} and includes a few variants, including Word2Vec and Binary2Img. All Instruction2Vec related methods use the assembly instructions as input and do not consider the structural information.

\textbf{FFmpeg baselines} For the FFmpeg dataset, we compared DeepEXE to Devign~\cite{zhou2019devign}, which provides the source code for FFmpeg. Devign uses a gated graph recurrent network (GGRN)~\cite{ruiz2020gated} as a graph learning technique. Unlike binary code, where only CFG can be extracted, Devign detects vulnearbilities at the source code level. It takes several intermediate graph representations of source code, such as AST, CFG, DFG, and NCS. We include several variants of Devign, such as Bi-LSTM, GGRN with CFG or all graphs, and Devign with CFG or all graphs. Note that we directly compare our results on the binary code with the original results of Devign, which are based on source code. 

\textbf{Esh baselines} To the best of our knowledge, Esh is not used in any other papers for vulnerability detection evaluation. The original paper that provided this dataset evaluates it at the basic block level and focuses on code matching. Therefore, we compare DeepEXE to the Bi-LSTM and GCN baselines we implemented ourselves.

\subsection{Preliminaries}
\label{appendix: preliminaries}
\textbf{Graph Neural Network} GNN is a topological learning technique for input data with graph structures. A graph is represented as $\gG=(V,E)$ that contains $n:=|V|$ nodes and $e:=|E|$ edges. An edge $E_{ij}:=(V_i, V_j)$ represents the directed or un-directed connection between node $(i,j)$. In practice, the edge information is represented in the form of an adjacency matrix $\mA\in\mathbb{R}^{n\times n}$. Generally, one can obtain some initial node embedding $U\in\mathbb{R}^{n\times h}$ before feeding into the network. The message passing (i.e. node aggregation) is performed at each GNN layer as follows:
\begin{equation}
    X^{t+1} = \phi(X^{t}W^t\mA^{t}) 
\end{equation}
where $W^t \in \mathbb{R}^{h\times h}$ is a trainable parameter at layer $t$. Each message passing step aggregates 1-hop neighbour information into the current node given that an edge exists in $\mA$. The final node vector $X^T$ then learns the topological information from all $T$-hop away neighbours. In case of graph classification, a pooling layer such as add pooling can be used to obtain the graph embedding $\mG$:
\begin{equation}
    \mG = \sum_i^nX^T_{i,j}, \forall~j=1,...,h
\end{equation}

\textbf{REINFORCE Algorithm} Reinforcement learning is a class of algorithms that specify the actions within an environment that optimizes the reward $r$. In particular, the REINFORCE algorithm~\cite{williams1992simple} is a form of policy gradient algorithm that computes the stochastic gradient with respect to the reward. It involves a state $s$ that can be obtained from a neural network, an agent $a$ that specifies the action space $\mathcal{A}$, and a policy $\pi(a|s)$ that takes the action $a$ given a state $s$ with probabilities. Usually, the policy is randomly initialized and the algorithm iterates through epochs, where backpropagation is performed at each epoch to update the policy in the context of a neural network setup.

\subsection{Well-posedness} 
Equation~(\ref{eqn:implicit}) needs to have a unique solution $X^*$ when iterated infinitely. Such property is called the well-posedness. According to Gu et al.~\cite{gu2020implicit}, $W$ and $\Tilde{\mA}$ are well-posed for $\phi$ when there is a unique solution. First of all, the choice of $\phi$ needs to satisfy the component-wise non-expansive (CONE) property, where most activation functions such as ReLU, Sigmoid, and Tanh, possess such property~\cite{el2021implicit}. Then, we need to construct sufficient conditions on $W$ and $\Tilde{\mA}$ with a CONE activation function for well-posedness. It is stated that $||W||_\infty < \kappa/\lambda_{pf}(\Tilde{\mA})$ needs to be true, where $||W||_\infty$ is the infinity norm, $\lambda_{pf}(\Tilde{\mA})$ is the Perron-Frobenius (PF) eigenvalue~\cite{berman1994nonnegative}, $\kappa \in [0,1)$ is the scaling constant. Equation~(\ref{eqn:implicit}) then has a unique solution. This is ensured by projecting $W$ in each update to satisfy this condition:
\begin{equation}
    W' = \argmin_{||M||_{\infty}\leq\kappa/\lambda_{pf}(\Tilde{\mA})}||M-W||^2_F
\end{equation}
where $||\cdot||_F$ is the Frobenius norm. Note that even with a gated convolution which results in an updated $\Tilde{\mA}$ for every iteration, we still maintain a well-posed $\Tilde{\mA}$ as it contains a strictly smaller or equal PF eigenvalue than the original $A$, given the agent $a$ is non-expansive, resulting in $\kappa/\lambda_{pf}(\Tilde{\mA}) \geq \kappa/\lambda_{pf}(A)$.

\end{document}